\documentclass[sigconf]{acmart}
\usepackage{algorithm}
\usepackage{algpseudocode}
\usepackage{multirow}
\usepackage{tikz}
\usetikzlibrary{positioning}
\usepackage{pgfplots}
\usepackage{subcaption}

\usepackage{booktabs} 

\setcopyright{rightsretained}

\acmConference[DAC'17]{Design Automation Conference}{2017}{San Francisco, CA} 
\acmYear{2017}
\copyrightyear{2017}

\algdef{SE}[PROCEDURE]{Procedure}{EndProcedure}%
   [2]{\algorithmicprocedure\ \textproc{#1}\ifthenelse{\equal{#2}{}}{}{(#2)}}%
   {\algorithmicend\ \algorithmicprocedure}%

\begin{document}
\title{Cryptographically Secure Multi-Tenant Provisioning of FPGAs}
\subtitle{}

\author{Arnab Bag, Sikhar Patranabis, Debapriya Basu Roy and Debdeep Mukhopadhyay}
\affiliation{\institution{Indian Institute of Technology Kharagpur}}
\email{arnabbag@iitkgp.ac.in,sikhar.patranabis@iitkgp.ac.in, deb.basu.roy@cse.iitkgp.ernet.in, debdeep$@cse.iitkgp.ernet.in }

\begin{abstract}

FPGAs~(Field Programmable Gate arrays) have gained massive popularity today as accelerators for a variety of workloads, including big data analytics, and parallel and distributed computing. This has fueled the study of mechanisms to provision FPGAs among multiple tenants as general purpose computing resources on the cloud. Such mechanisms offer new challenges, such as ensuring IP protection and bitstream confidentiality for mutually distrusting clients sharing the same FPGA. A direct adoption of existing IP protection techniques from the single tenancy setting do not completely address these challenges, and are also not scalable enough for practical deployment. In this paper, we propose a dedicated and scalable framework for secure multi-tenant FPGA provisioning that can be easily integrated into existing cloud-based infrastructures such as OpenStack. Our technique has \emph{constant resource/memory overhead} irrespective of the number of tenants sharing  a given FPGA, and is provably secure under well-studied cryptographic assumptions. A prototype implementation of our proposition on Xilinx Virtex-7 UltraScale FPGAs is presented to validate its overheads and scalability when supporting multiple tenants and workloads. To the best of our knowledge, this is the first FPGA provisioning framework to be prototyped that achieves a desirable balance between security and scalability in the multi-tenancy setting.

\end{abstract}

\keywords{FPGAs, Security, Provisioning, Multi-Tenant, Cloud Computing}

\maketitle

\section{Introduction}

The modern era of cloud computing has actualized the idea of ubiquitous provisioning of computational resources and services via a network. Cloud-based solutions are now marketed by all leading enterprise IT vendors such as IBM~(PureApplication), Oracle~(ExaData), Cisco~(UCS) and Microsoft~(Azure), as well as Web companies such as Amazon~(AWS) and Google~(Compute Engine). In the midst of this paradigm shift from traditional IT infrastructures to the cloud, FPGAs (Field Programmable Gate Arrays) have risen as attractive computational avenues for accelerating heavy workloads. Modern FPGAs offer a number of advantages including, but not limited to, reconfigurability, high throughput, predictable latency and low power consumption. They also offer \emph{dynamic partial reconfiguration}~(DPR) capabilities \cite{lie2009dynamic}, that allow non-invasive run-time modification of existing circuitry for on-the-fly functionality enhancement. This is particularly beneficial when a given FPGA is shared simultaneously by multiple tenants: \emph{an individual tenant can re-configure her share of the FPGA resources at any time, without disturbing the applications being run by other tenants}. There is, in fact, a growing demand today for deploying FPGAs as general purpose computing resources on the cloud.

\noindent\textbf{Security Challenges.} Provisioning FPGAs on the cloud offers a number of challenges such as resource abstraction, ecosystem compatibility (libraries and SDKs) and, most importantly, \emph{security}. While some of these challenges have been addressed comprehensively in the existing literature~\cite{chen2014enabling}, security issues emerging from such a model are largely under-studied. One such security issue is \emph{IP protection}. Multiple mutually distrusting tenants sharing a common pool of FPGA resources are likely to demand guarantees for bitstream confidentiality. Since FPGAs are inherently designed for single party access, FPGA vendors today focus on ensuring the privacy of bitstreams originating from \emph{single users}, especially when deployed into hostile industrial/military environments. Mitigation techniques typically used bitstream encryption and authentication, combined with fault-tolerance. However, a direct adoption of such techniques in the multi-tenancy setting potentially blows up resource-requirements, imposes significant key-management overheads, and leads to an overall lack of scalability. This motivates the need for dedicated and scalable security solutions tuned to the multi-tenancy setting.

\noindent\textbf{Existing Solutions.} While a number of recent works~\cite{so2008unified,kirchgessner2012virtualrc,opitz2012accelerating} have helped develop general acceptance for FPGAs as general-purpose computing elements in portable ecosystems, security concerns regarding large-scale FPGA deployment been discussed only in the context of specific applications. For example, the authors of \cite{arasu2013secure} have looked into the security of specific applications such as building databases, where FPGAs are used as accelerators. Their security discussions are more at the application-level rather than the system-level. Other works~\cite{chen2014enabling} focus on the threats originating from malicious tenants either crashing the system or attempting illegal memory accesses. Their proposed mitigations are mostly based on virtualization, in the sense that they use dedicated hypervisors and DMA units to regulate the memory access made by each tenant's bitstream file on the host FPGA node. However, they do not consider the threats posed by \emph{co-resident VM attacks} \cite{ristenpart2009hey,irazoqui2014wait}, where data resident on a target VM can be stolen by a second malicious VM, so long as they co-exist on the same physical node. This poses a massive threat to IP security in the shared tenancy setting, and underlines the need for cryptographic security guarantees in addition to architectural barricading. While a number of cryptographic solutions have been proposed for IP protection in the single tenancy scenario~\cite{kean2002cryptographic,guajardo2007fpga}, there exist no equivalent solutions tuned to the shared tenancy setting to the best of our knowledge.

\noindent\textbf{Our Proposition.} In this paper, we propose a dedicated and scalable framework for secure multi-tenant FPGA provisioning on the cloud. Our framework also has the following desirable features: 
\begin{itemize}
\item Our framework guarantees bitstream confidentiality in exchange for a constant amount of resource/memory overhead, \emph{irrespective of the number of tenants sharing  a given FPGA}. We achieve this using a novel technique known as \emph{key-aggregation} that is provably secure under well-studied cryptographic assumptions. 
\item  The only trusted agent in our framework is the FPGA vendor. Note that even in IP protection solutions in the single tenancy setting, the FPGA vendor is typically a trusted entity. Hence, this is a reasonable assumption. More importantly, the cloud service provider need not be trusted, which is desirable from a tenant's point of view.
\item Our framework can be easily integrated into existing cloud-based infrastructures such as OpenStack, and does not interfere with other desirable properties of an FPGA provisioning mechanism, such as resource virtualization/isolation and platform compatibility.
\end{itemize}
 \noindent\textbf{Prototype Implementation.} We illustrate the scalability of our proposed approach via a prototype implementation on Xilinx Virtex-7 UltraScale FPGAs. Our results indicate that the proposed approach has a fixed overhead of around $10-15\%$ of the available FPGA resources. This overhead remains unaltered for any number of tenants/workloads using the FPGA resources at any given point of time. To the best of our knowledge, this is the first FPGA provisioning framework to be prototyped that achieves a desirable balance between security and scalability in the multi-tenancy setting.

\noindent \textbf{Applications in the Automotive Setting.} FPGAs are being increasingly used as accelerators in automotive applications. In particular, the high parallel processing capabilities of FPGAs provide great advantages in applications such as ADAS, Smart Park Assist systems, and power control systems in modern vehicles. Most FPGAs also come with integrated peripheral cores that implement commonly-used functions like communication over controller area network~(CAN)~\cite{johansson2005vehicle}. In an automotive setting, a single FPGA may be required to accelerate applications from multiple stakeholders, that are mutually distrusting and wish to protect their individual IPs. The core techniques underlying our proposed framework in this paper can be equivalently applied to build efficient and scalable IP protection units for such applications. 

\section{Secure Multi-Tenant FPGA Provisioning: Our Proposition}

\begin{figure}[t]
	\centering
	\includegraphics[scale=0.33]{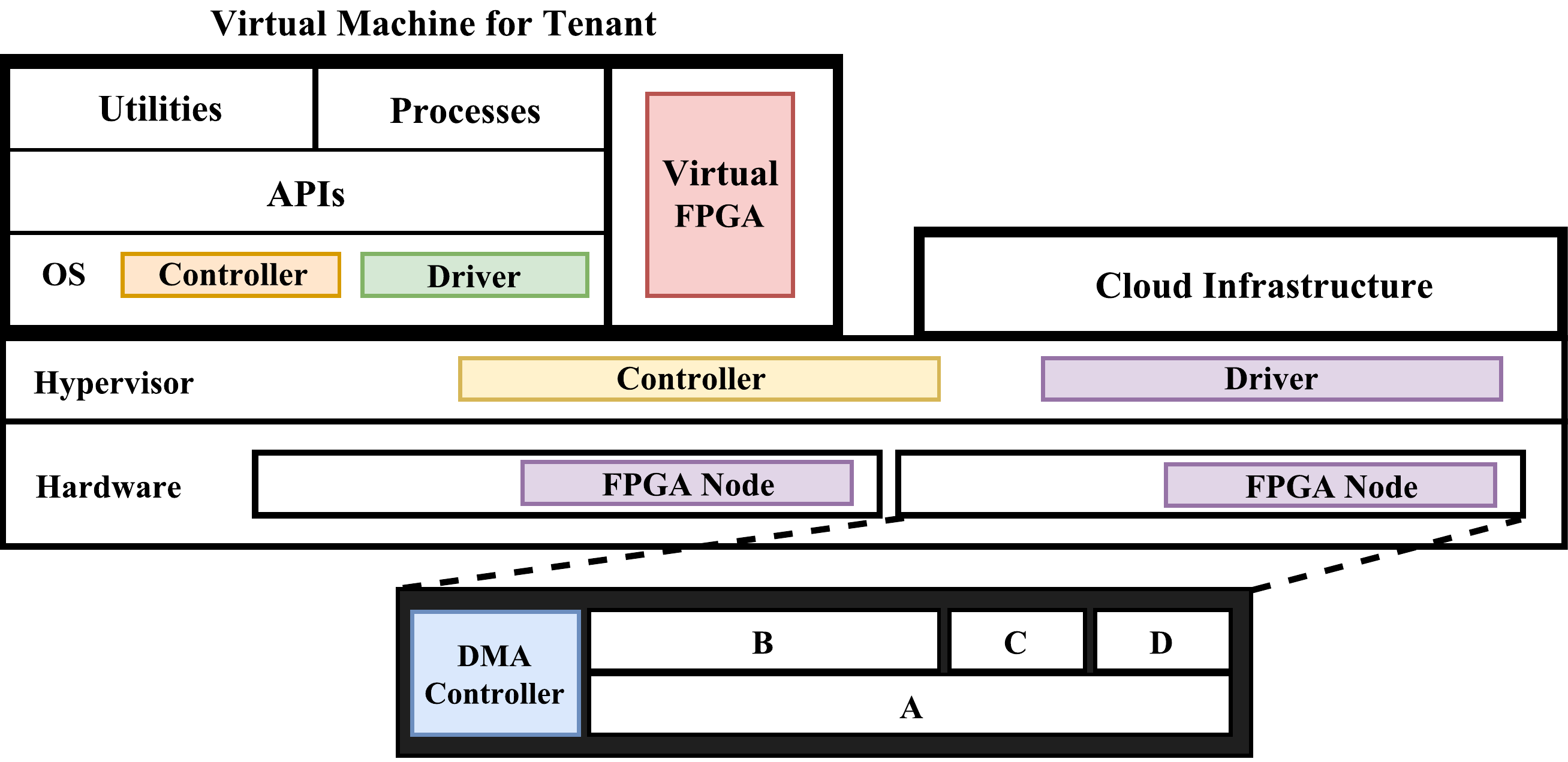}
    \caption{FPGA Provisioning on a Cloud~\cite{chen2014enabling}}
    \label{fig:overview}
\end{figure}

In this section, we present our proposal for secure provisioning of FPGAs among multiple tenants on the cloud. We assume a basic  FPGA provisioning setup on a cloud~\cite{chen2014enabling}, as illustrated in Figure \ref{fig:overview}. The idea is to abstract the FPGA resources to the client as an \emph{accelerator pool}. Each FPGA is divided into multiple slots (e.g. A, B, C and D in Figure \ref{fig:overview}), with one or more slots assigned to a tenant. The dynamic partial reconfiguration mechanism of modern FPGAs allows a tenant to view each such slot as a \emph{virtual FPGA}, with specific resource types, available capacity and compatible interfaces. The DMA controller module is meant primarily for bandwidth and priority management across the various FPGA partitions. At the hypervisor layer, the controller module chooses available FPGA nodes based on their compatibility with a tenant's requirements, and helps configure them with the desired bitstream file via the service layer. The tenant essentially sees a VM, embedded with a virtual FPGA and containing the necessary APIs and controller modules to configure the FPGA. The allocation of resources to various tenants and the creation of corresponding VMs is handled by a separate controller module. More details of this basic setup can be found in~\cite{chen2014enabling}. Our aim is to propose an efficient and secure mechanism that ensures IP protection in this setup, without compromising on the other well-established features such as virtualization, inter-VM isolation and platform compatibility.

\subsection{Bring Your Own Keys~(BYOK)}

The fundamental idea underlying our security proposal is as follows: each tenant encrypts her bitstream using a secret-key \emph{of her own choice} before configuring the virtual FPGA with the same. Since bitstreams would potentially be encrypted in bulk, a symmetric-key encryption algorithm such as AES-128 is the ideal choice in this regard. Note that this approach immediately assures bitstream confidentiality. In particular, since neither the service provider nor any malicious agent can correctly guess the key chosen by a tenant (except with negligible probability), they can no longer gain access to her bitstream.

Notwithstanding its apparent benefits, the aforementioned BYOK-based bitstream encryption technique poses two major challenges in the shared FPGA setting - synchronizing bitstream encryption and decryption for different tenants, and efficient key-management. The main novelty of our proposal is in the application of \emph{key-aggregation}~\cite{patranabis2017provably} - a provably secure cryptographic technique - to efficiently solve both these challenges. We begin by providing a brief overview of a key-aggregate cryptosystem~(KAC), along with a concrete construction for the same. We then demonstrate how KAC solves the key-management and synchronization challenges posed by the BYOK-based approach.\\

\subsection{Key-Aggregate Cryptosystems~(KAC)}

\begin{figure}[t]
	\centering
	\includegraphics[scale=0.3]{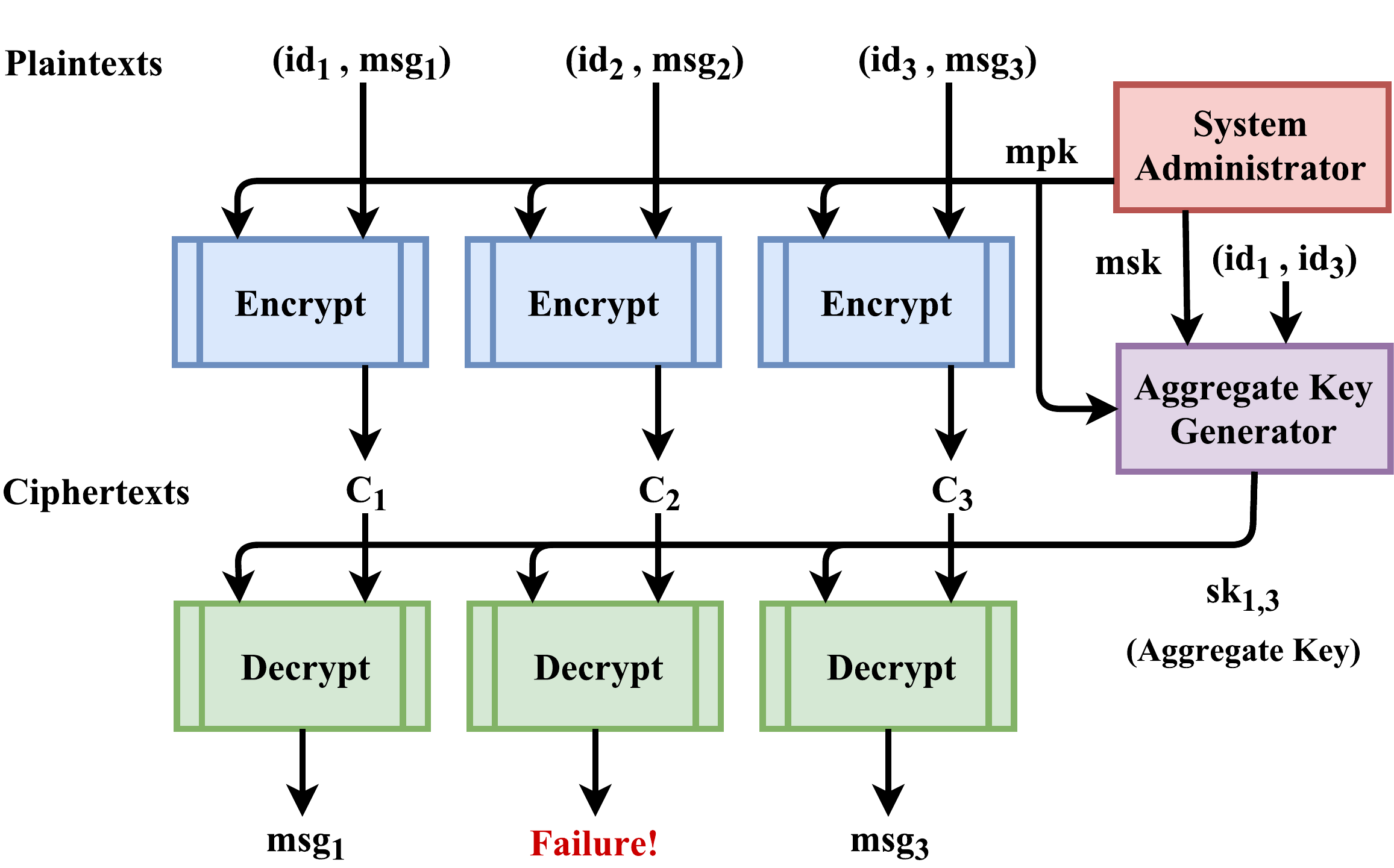}
    \caption{An Illustration of KAC over Three Entities}
    \label{fig:KAC}
\end{figure}

KAC is a public-key mechanism to encapsulate multiple decryption-keys corresponding to an arbitrarily large number of independently encrypted entities into a single constant-sized entity. In a KAC, each plaintext message/entity is associated with a unique identity $\text{\textsf{id}}$, and is encrypted using a common master public-key $\text{\textsf{mpk}}$, generated by the system administrator. The system administrator also generates a master secret-key $\text{\textsf{msk}}$, which in turn is used to generate decryption keys for various entities. The main advantage of KAC is its ability to generate constant-size \emph{aggregate decryption keys}, that combine the power of several individual decryption keys. In other words, given ciphertexts $C_1,C_2\cdots,C_n$ corresponding to identities $\text{\textsf{id}}_1,\text{\textsf{id}}_2,\cdots,\text{\textsf{id}}_n$, it is possible to generate a constant-size aggregate decryption key $\text{\textsf{sk}}_{\mathcal{S}}$ for any arbitrary subset of identities $\mathcal{S}\subseteq\{\text{\textsf{id}}_1,\cdots,\text{\textsf{id}}_n\}$. In addition, the aggregate key $\text{\textsf{sk}}_{\mathcal{S}}$ \emph{cannot} be used to decrypt any ciphertext $C_j$ corresponding to an identity $\text{\textsf{id}}_j\notin\mathcal{S}$. Figure \ref{fig:KAC} illustrates the concept of a KAC scheme with a simple toy example. Observe that the individual secret-keys $\text{\textsf{sk}}_{1}$ and $\text{\textsf{sk}}_{3}$ for the identities $\text{\textsf{id}}_1$ and $\text{\textsf{id}}_3$ are compressed into a single aggregate-key $\text{\textsf{sk}}_{1,3}$, that can be used to decrypt both the ciphertexts $C_1$ and $C_3$, \emph{but not} $C_2$. Additionally, $\text{\textsf{sk}}_{1,3}$ has the same size as either of $\text{\textsf{sk}}_{1}$ and $\text{\textsf{sk}}_{3}$, individually.

\noindent\textbf{A Concrete KAC Construction on Elliptic Curves.} Algorithm \ref{alg:KAC} briefly describes a provably secure construction for KAC to illustrate its key-aggregation property. The main mathematical structure used by the construction is a prime order sub-group of elliptic curve points $\mathbb{G}$, generated by a point $P$, and a bilinear map $e$ that maps pairs of elements in $\mathbb{G}$ to a unique element in another group $\mathbb{G}_T$. The construction supports a maximum of $n$ entities, and is provably secure against chosen-plaintext-attacks under a variant of the bilinear Diffie-Hellman assumption~\cite{DBLP:journals/joc/Joux04}. We refer the reader to~\cite{patranabis2017provably} for more details on the correctness and security of the construction. Note that the notations $P_1+P_2$  and $[a]P$ denote point addition and scalar multiplication operations, respectively, over all elliptic curve points $P,P_1,P_2$ and all scalars $a$. Observe that the aggregate key $\text{\textsf{sk}}_{\mathcal{S}}$ is a \emph{single elliptic-curve point} (with a fixed representation size), irrespective of the size of the subset $\mathcal{S}$.

\begin{algorithm}
\caption{A Concrete KAC construction on Elliptic Curves}
\label{alg:KAC}
 \begin{algorithmic}[1]
 \Procedure{$\text{\textsf{KAC.Setup}}$}{n}
 \State Take as input the number of entities $n$
 \State Let $P$ be an elliptic curve point of prime order $q$ that generates a group $\mathbb{G}$ with a bilinear map $e:\mathbb{G}\times\mathbb{G}\longrightarrow \mathbb{G}_T$.
 \State Randomly choose $\alpha,\gamma$ in the range $[0,q-1]$ and output the following:
 \begin{eqnarray}
 \text{\textsf{mpk}} &=& \left(\{\left[\alpha^j\right]P\}_{j\in[0,n]\cup[n+2,2n]},\left[\gamma\right] P\right)\nonumber\\
 \text{\textsf{msk}} &=& \gamma\nonumber
 \end{eqnarray}
 \EndProcedure
 \Procedure{$\text{\textsf{KAC.Encrypt}}$}{$\text{\textsf{mpk}},i,M$}
 \State Take as input the master public key $\text{\textsf{mpk}}$, an entity identity $i\in[1,n]$ and a plaintext bitstream $M$.
 \State Randomly choose $r$ in the range [0,q-1] and set:
 \begin{eqnarray}
 c_0 &=& \left[r\right]P\nonumber\\
 c_1 &=& \left[r\right]\left(\left[\gamma\right] P+\left[\alpha^i\right]P\right)\nonumber\\
 c_2 &=& M\oplus H\left(e\left(\left[\alpha^1\right]P,\left[\alpha^n\right]P\right)^r\right)\nonumber
 \end{eqnarray}
 \noindent where $H$ is a collision-resistant hash function and $\oplus$ denotes the bit-wise XOR operation
 \State Output the ciphertext $C = \left(c_0,c_1,c_2\right)$ 
 \EndProcedure
 \Procedure{$\text{\textsf{KAC.AggregateKey}}$}{$\text{\textsf{msk}},\text{\textsf{mpk}},\mathcal{S}$}
 \State Take as input the master secret key $\text{\textsf{msk}}=\gamma$, the master public key \text{\textsf{mpk}} and a subset of entities $\mathcal{S}\subseteq[1,n]$.
 \State Compute $a_{\mathcal{S}} = \sum_{j\in\mathcal{S}}\left[\alpha^{n+1-j}\right]P$
 \State Output the aggregate key $\text{\textsf{sk}}_{\mathcal{S}} = \left[\gamma\right]a_{\mathcal{S}}$
 \State Also output $a_{\mathcal{S}}$ and $b_{i,\mathcal{S}} = \sum_{j\in\mathcal{S}\setminus\{i\}}\left[\alpha^{n+1-j+i}\right]P$ for each $i\in\mathcal{S}$
 \EndProcedure
 \Procedure{$\text{\textsf{KAC.Decrypt}}$}{$\text{\textsf{sk}}_{\mathcal{S}},a_{\mathcal{S}},b_{i,\mathcal{S}},C$} 
 \State Take as input a ciphertext $C = \left(c_0,c_1,c_2\right)$ corresponding to an entity with identity $i$, an aggregate key $\text{\textsf{sk}}_{\mathcal{S}}$ such that $i\in\mathcal{S}$, along with $a_{\mathcal{S}}$ and $b_{i,\mathcal{S}}$ as defined above.
 \State Output the decrypted message $M$ as:
 \begin{equation}
 M = c_2\oplus H\left(e\left(a_{\mathcal{S}},c_1\right)\cdot e\left(\text{\textsf{sk}}_{\mathcal{S}}+b_{i,\mathcal{S}},c_0\right)^{-1}\right)\nonumber
 \end{equation}
 \noindent where $H$ is the same collision-resistant hash function as used in $\text{\textsf{KAC.Encrypt}}$
 \EndProcedure
 \end{algorithmic}
\end{algorithm}

\subsection{Combining BYOK with KAC}

\begin{figure}[t]
	\centering
	\includegraphics[scale=0.35]{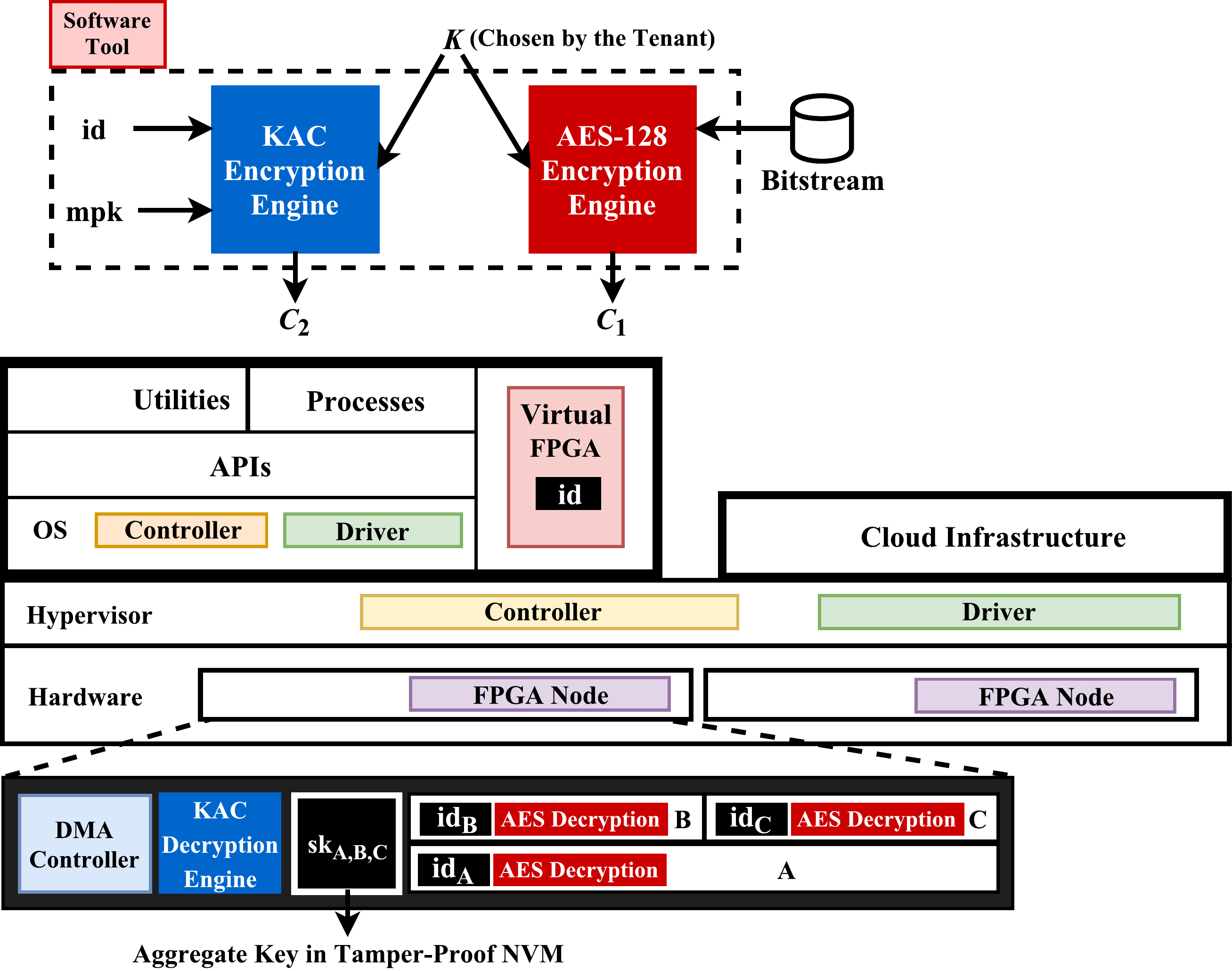}
    \caption{Secure FPGA Provisioning Scheme: Combining KAC with BYOK}
    \label{fig:final}
\end{figure}

\begin{algorithm}
\caption{Secure Multi-Tenant FPGA Provisioning}
\label{alg:proposal}
 \begin{algorithmic}[1]
	\Procedure{Initial Setup (FPGA Vendor)}{}     
    \State $\left(\text{\textsf{mpk}},\text{\textsf{msk}}\right)\leftarrow\text{\textsf{KAC.Setup}}$
     \State Publish the master public key $\text{\textsf{mpk}}$
     \For{each manufactured FPGA}
     	\For{each FPGA partition}
     		\State Assign a unique random identity $\text{\textsf{id}}$ to the partition
     	\EndFor 
        \State Let $\mathcal{S}$ denote the set of all $\text{\textsf{id}}$-s corresponding to partitions on the same FPGA
        \State $\left(\text{\textsf{sk}}_{\mathcal{S}},a_{\mathcal{S}},\{b_{\text{\textsf{id}},\mathcal{S}}\}_{\text{\textsf{id}}\in\mathcal{S}}\right)\leftarrow\text{\textsf{KAC.AggregateKey}}\left(\text{\textsf{msk}},\mathcal{S}\right)$
        \State Embed $\text{\textsf{sk}}_{\mathcal{S}}$ in a tamper-proof non-volatile memory segment on the FPGA
        \State Embed $a_{\mathcal{S}}$ in a non-volatile memory segment on the FPGA (\emph{need not be secure/tamper-proof}). 
        \State Embed each $b_{\text{\textsf{id}},\mathcal{S}}$ in a non-volatile memory segment of the partition with identity \text{\textsf{id}} (\emph{again need not be secure/tamper proof})
     \EndFor      
     \State Each FPGA is provisioned with a single KAC decryption engine, while each FPGA partition is provisioned with its own AES-128 decryption engine.
     \EndProcedure
	 \Procedure{Bitstream Encryption}{\text{\textsf{Bitstream}}}
     \State Suppose a tenant is assigned an FPGA partition with identity $\text{\textsf{id}}$.
     \State $K\leftarrow\text{\textsf{AES.KeyGen}}$
     \State $C_1\leftarrow \text{\textsf{AES.Encrypt}}\left(K,\text{\textsf{Bitstream}}\right)$
     \State $C_2\leftarrow\text{\textsf{KAC.Encrypt}}\left(\text{\textsf{mpk}},\text{\textsf{id}},K\right)$
     \State Submit $\left(C_1,C_2\right)$ to the framework for configuring the FPGA partition.
     \EndProcedure
     \Procedure{Bitstream Decryption}{$C_1,C_2$} 
     \State $K\leftarrow\text{\textsf{KAC.Decrypt}}\left(\text{\textsf{sk}}_{\mathcal{S}}, C_2\right)$
     \State $\text{\textsf{Bitstream}}\leftarrow \text{\textsf{AES.Decrypt}}\left(K,C_1\right)$
     \EndProcedure
 \end{algorithmic}
\end{algorithm}

The crux of our proposal lies in combining BYOK with KAC for efficient key-management and synchronization of bitstream encryption-decryption. We achieve this via the following three-step proposal:

\noindent\textbf{Step-1: Setup.} In this step, the FPGA vendor sets up a KAC system by generating a master public key and a master secret key. Each manufactured FPGA can be divided into a maximum of $n$ partitions, where each partition is associated with a unique partition identity $\text{\textsf{id}}$, and represents an independent virtual FPGA from the tenant point of view. Each FPGA contains a KAC decryption engine, that is pre-programmed to use a single aggregate decryption key $sk_{\mathcal{S}}$ corresponding to the subset $\mathcal{S}$ of partition ids it hosts. In a Xilinx Virtex-7 FPGA, the aggregate key can be securely stored in either a dedicated non-volatile RAM (often backed up by a small externally connected battery), or in the eFUSE~\footnote{\url{https://www.xilinx.com/support/documentation/application_notes/xapp1239-fpga-bitstream-encryption.pdf}}.

\noindent\textbf{Step-2: Bitstream Encryption.} In keeping with the idea behind BYOK, each tenant encrypts her bitstream using her own AES-128 key. Commercially available software tools such as Xilinx Vivado already provide such facilities. We simply propose augmenting this functionality to additionally encrypt the AES-128 key using the master public key of the KAC. The second encryption is performed under the identity $\text{\textsf{id}}$ of the partition assigned to the tenant.

\noindent\textbf{Step-3: Bitstream Decryption.} Bitstream encryption occurs on-chip in two steps. Each FPGA is provided with a single KAC decryption core, while each individual partition is provided with its own AES-128 decryption core. The KAC decryption engine is first used to recover the AES-128 key chosen by the tenant. Since a single tenant is expected to use the same AES-128 key in a given session, the KAC decryption core needs to be invoked only once per tenant. The recovered key is subsequently used to decrypt any number of encrypted bitstreams and program the FPGA partition with the same. 

\noindent Quite evidently, the proposal has the following desirable features from the point of view of efficiency as well as security:
\begin{itemize}
\item \textbf{Constant Secure Storage Overhead per FPGA:} Each FPGA stores a single aggregate decryption key that suffices for all its partitions. As already mentioned, KAC generates constant-overhead aggregate-keys irrespective of the number of entities they correspond to. Hence, the memory requirement per FPGA for secure key storage remains the same irrespective of the maximum number of partitions $n$. In other words, the framework scales to any arbitrarily large $n$ without incurring any additional overhead for secure key storage.
\item\textbf{Constant Encryption and Decryption Latency:} The encryption and decryption latencies for both KAC and AES-128 are constant, and independent of the maximum number of partitions $n$ supported by an FPGA. In particular, the encryption and decryption sub-routines in Algorithm~\ref{alg:KAC} involve a constant number of elliptic curve operations, and hence require a constant amount of time.
\item \textbf{No Leakage to the Cloud Service Provider:} The new scheme achieves synchronization between the encryption and decryption engines via a public-key mechanism that is set up by the FPGA vendor. Since the entire bitstream decryption happens on-chip, the confidentiality of the bitstream as well as that of the AES-128 key from the cloud service provider (as well as any external malicious agents) are guaranteed by the security of AES-128 and the CPA security of the KAC scheme, respectively. 
\end{itemize}

\section{Prototype Implementation}

In this section, we present a prototype implementation for the secure FPGA provisioning framework described in the previous section. In particular, we focus on the overhead and performance results for the security-related components, namely KAC and AES-128. The results are presented in two parts. The first part focuses on the on-chip decryption engines, while the second part focuses on the software tool for generating the encrypted bitstreams and encrypted AES-128 keys.

\begin{table*}[t]
\centering
\caption{Implementation Details: Core Arithmetic Blocks}
\label{tab:cores}
\vspace*{-3mm}
\scalebox{0.8}{
\begin{tabular}{|c|c|c|c|c|c|}
 \hline\hline 
 \multirow{4}{*}{\textbf{Core blocks over $F_p$}} & \textbf{Module} & \textbf{LUT Count} & \textbf{Register Count} & \textbf{DSP Blocks} & \textbf{Latency (in ms)}\\\cline{2-6}
 & Adder & 686 & 662 & 0 & $4 \times 10^{-5}$  \\\cline{2-6}
 & Multiplier & 5571 & 2943 & 11 & $1.465 \times 10^{-3}$ \\\cline{2-6}
 & Inverter & 5907 & 3838 & 11 & 0.375 \\\hline\hline
\end{tabular}}
\end{table*}

\begin{table*}[t]
\centering
\caption{Implementation Details: Elliptic Curve Operations and Tate Pairing}
\label{tab:elliptic}
\vspace*{-3mm}
\scalebox{0.8}{
\begin{tabular}{|c|c|c|c|c|}
 \hline\hline 
 \multirow{3}{*}{\textbf{Elliptic Curve Operations}} & \textbf{Module} & \textbf{Clock Cycles} & \textbf{Operating Frequency(MHz)} & \textbf{Latency (in ms)}\\\cline{2-5}
 & Point Addition & 6231 & \multirow{2}{*}{200}& $3.116 \times 10^{-2}$ \\\cline{2-3}\cline{5-5}
 & Point Doubling & 5330 & & $2.650 \times 10^{-2}$ \\\hline\hline
 \multirow{3}{*}{\textbf{Tate Pairing Operations}} & \textbf{Module} & \textbf{Clock Cycles} & \textbf{Operating Frequency(MHz)} & \textbf{Latency (in ms)}\\\cline{2-5}
 & Miller's Algorithm & 83403777& \multirow{2}{*}{200} & 417.018 \\\cline{2-3}\cline{5-5}
 & Final Exponentiation & 96829996 & & 483.296 \\\hline\hline
\end{tabular}}
\end{table*}

\begin{table*}[t]
\centering
\caption{Implementation Details: On-Chip Decryption Engines}
\label{tab:decrypt}
\vspace*{-3mm}
\scalebox{0.8}{
\begin{tabular}{|c|c|c|c|c|c|c|c|c|c|c|}
 \hline \hline
\multirow{2}{*}{\textbf{Decryption Module}} & \multicolumn{3}{|c|}{\textbf{Resources Consumed}} & \multicolumn{3}{|c|}{\textbf{Overall Resources(Virtex-7 FPGA)}} & \multicolumn{3}{|c|}{\textbf{Percentage Consumption}} & \multirow{2}{*}{\textbf{Latency(ms)}}\\\cline{2-10}
& LUTs & Registers & DSP Blocks & LUTs & Registers & DSP Blocks & LUTs & Registers & DSP Blocks & \\\hline
KAC & 78196 & 46828 & 99 & \multirow{2}{*}{607400}& \multirow{2}{*}{866400} & \multirow{2}{*}{3600}    & $12.87\%$ & $5.40\%$ & $2.75\%$ & 1802.332\\\cline{1-4}\cline{8-11}
AES-128 & 1484 & 149 & 0 & & &    & $2.44\times 10^{-1}\%$ & $1.72\times 10^{-2}\%$ & $0\%$ & $8 \times 10^{-5}$\\\hline\hline
\end{tabular}}
\end{table*}

\begin{table}[t]
\centering
\caption{Implementation Details: KAC Encryption Engine}
\label{tab:encrypt}
\vspace*{-3mm}
\scalebox{0.8}{
\begin{tabular}{|c|c|c|c|c|}
 \hline
\textbf{Operation} & Point Addition & Point Doubling & Tate Pairing & KAC Encryption\\\hline
 \textbf{Latency(ms)} & $3.041\times 10^{-2}$ &  $1.568\times 10^{-2}$ & 1172.413 & 1176.981\\\hline
\end{tabular}}
\vspace*{-7mm}
\end{table}

\subsection{On-Chip Decryption Engines}

We implemented the decryption engines for KAC and AES-128 on a Virtex-7 UltraScale FPGA. In this section, we present post-placement and routing results to illustrate their overhead and operational latencies. To implement the KAC decryption engine, we chose an elliptic curve that offers a a 128-bit security level from the family of pairing-friendly Barreto-Naehrig (BN) curves~\cite{DBLP:conf/sacrypt/BarretoN05}. The curve and all associated operations (point addition and doubling) are defined over a finite field $F_p$ (where $p$ is a $256$-bit prime). On this curve, we implemented the well-known bilinear Tate pairing operation~\cite{frey1999tate}, which in turn uses Miller's algorithm~\cite{miller2004weil} followed by a final exponentiation~\cite{galbraith2002implementing}. The group order $q$ for the pairing operation is a $128$-bit prime factor of $p^{12}-1$. The Miller's algorithm operates over $F_p$ and the quadratic extension field $F_{p^2}$, and runs for $\log_2{q}$ many iterations. The final exponentiation is performed in the extension field $F_{p^{12}}$, and raises the output of Miller's algorithm to the power $\left(p^{12}-1\right)/q$. Additional mathematical details related to the Tate pairing algorithm can be found in~\cite{frey1999tate,galbraith2002implementing}. Note that while alternative elliptic curves with smaller characteristics primes (e.g. $p=2$ or $p=3$) afford more hardware-efficient pairing implementations~\cite{duursma2003tate,kerins2005efficient,oliveira2007tinytate}, the security guarantees provided by such curves are presently under threat due to recent advances in DLP~\cite{DBLP:journals/dcc/JouxP16}. Finally, for the hash function $H$ in Algorithm~\ref{alg:KAC}, we use an FPGA-based implementation of SHA-256~\cite{chaves2006improving}.

\noindent\textbf{Multipliers using DSP Blocks.} A novel feature of our Tate pairing implementation as compared to existing work~\cite{ghosh2013secure} is the use of DSP blocks to design efficient multipliers over the field $F_p$. Modern FPGAs such as the Xilinx Virtex-7 UltraScale are inherently equipped with numerous DSP blocks, which can be used to design low-latency circuits for arithmetic operations. We exploited this fact to design a high-speed $F_p$ multiplier, that optimally uses these DSP blocks based on an efficient tiling algorithm~\cite{roy2014tile} for operand decomposition.

\noindent\textbf{Hardware Implementation Results.}  The post-route area and timing reports for the arithmetic cores over $F_p$ is presented in Table~\ref{tab:cores}. The post-route timing reports for the elliptic curve operations (point addition and point doubling) as well as the Tate pairing implementation are summarized in Table~\ref{tab:elliptic}. Table~\ref{tab:decrypt} summarizes the overall area and timing reports for the KAC and AES-128 decryption engines. As depicted in Algorithm~\ref{alg:KAC}, the decryption algorithm uses the Tate pairing core twice to compute the two pairings, followed by an application of the $F_{p^{12}}$ multiplication core to compute their product. Finally, the SHA-256 module is used to hash the output of this multiplication core, and recover the bitstream. To optimize area requirements, multiple operations using the same FPGA module are performed serially.


\subsection{Software Encryption Engine}

The software encryption engine in our prototype implementation allows a tenant to encrypt her bitstream using an AES-128 key of her own choice, and subsequently, encrypt this key under the KAC scheme. As mentioned previously, BYOK-based bitstream encryption can be readily availed using commercial design tools such as Xilinx Vivado. We implemented the KAC encryption engine in software using the open-source Pairing-Based Cryptography~(PBC) library~\footnote{\url{https://crypto.stanford.edu/pbc/}}, that provides APIs to compute Tate pairings over the BN family of elliptic curves. The only pre-requisite for using the PBC library is the open-source GNU Multiple Precision Arithmetic Library~\footnote{\url{https://gmplib.org/}}. The PBC library works on a variety of operating systems, including Linux, Mac OS, and Windows (32 and 64 bits). We present implementation results for the KAC encryption engine using the PBC library in Table~\ref{tab:encrypt}. The target platform is a standard desktop computer, with an Intel Core i5-4570 CPU, 3.8 Gb RAM, and an operating frequency of $3.20$GHz. It is important to note that similar to the decryption operation, the latency for KAC encryption is also independent of the number of partitions a given FPGA can support.

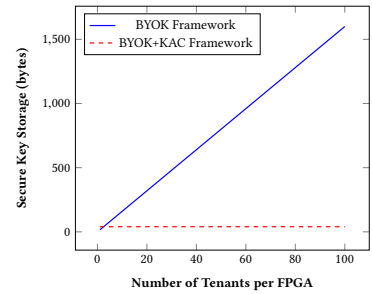
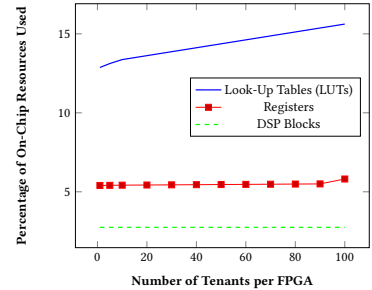
\begin{figure}[t]
\centering
\begin{subfigure}[Subfigure a]{0.3\textwidth}
\begin{tikzpicture}[scale = 0.57]
	 \begin{axis}[
	 		xlabel= \textbf{Number of Tenants per FPGA},
	 		ylabel= \textbf{Secure Key Storage (bytes)},
	 		legend pos={north west}
	 		]
	 \addplot[blue,thick] plot coordinates{
        (1, 16)
        (5, 80)
	 	(10, 160)
        (20, 320)
		(30, 480)
		(40, 640)
		(50, 800)
		(60, 960)
		(70, 1120)
        (80, 1280)
        (90, 1440)
        (100, 1600)
	 };\addlegendentry{BYOK Framework}
     
     \addplot[red,dashed,thick] plot coordinates{
	 	(1, 40)
        (5, 40)
	 	(10, 40)
        (20, 40)
		(30, 40)
		(40, 40)
		(50, 40)
		(60, 40)
		(70, 40)
        (80, 40)
        (90, 40)
        (100, 40)
	 };\addlegendentry{BYOK+KAC Framework} 
      
\end{axis}
\end{tikzpicture}
\caption{On-Chip Secure-Storage Requirements}
\label{plot:storage}
\end{subfigure}
\begin{subfigure}[Subfigure b]{0.3\textwidth}
\begin{tikzpicture}[scale = 0.57]
	 \begin{axis}[
	 		xlabel= \textbf{Number of Tenants per FPGA},
	 		ylabel= \textbf{Percentage of On-Chip Resources Used},
            legend style={at={(axis cs:37.5,8.5)},anchor=south west}
	 		]
	 \addplot[blue,thick] plot coordinates{
	 	(1, 12.87)
        (5, 13.12)
	 	(10, 13.37)
        (20, 13.62)
		(30, 13.87)
		(40, 14.12)
		(50, 14.37)
		(60, 14.62)
		(70, 14.87)
        (80, 15.12)
        (90, 15.37)
        (100, 15.62)
	 };\addlegendentry{Look-Up Tables~(LUTs)}
     
     \addplot plot coordinates{
	 	(1, 5.4)
        (5, 5.41)
	 	(10, 5.42)
        (20, 5.43)
		(30, 5.44)
		(40, 5.45)
		(50, 5.46)
		(60, 5.47)
		(70, 5.48)
        (80, 5.49)
        (90, 5.50)
        (100, 5.81)
	 };\addlegendentry{Registers}
     
     \addplot[green,dashed,thick] plot coordinates{
	 	(1, 2.75)
        (5, 2.75)
	 	(10, 2.75)
        (20, 2.75)
		(30, 2.75)
		(40, 2.75)
		(50, 2.75)
		(60, 2.75)
		(70, 2.75)
        (80, 2.75)
        (90, 2.75)
        (100, 2.75)
	 };\addlegendentry{DSP Blocks} 
\end{axis}
\end{tikzpicture}
\caption{On-Chip Resource Requirements}
\label{plot:area}
\end{subfigure}
\caption{Scalability of Our Proposed Framework }
\vspace*{-6mm}
\end{figure}

\section{Scalability of Our Framework}

In order to elucidate the scalability of our proposed framework, we demonstrate how the following parameters of our prototype implementation scale with the maximum number of tenants/partitions per FPGA:
\begin{itemize}
\item \textbf{Secure Storage on FPGA:} In Figure~\ref{plot:storage}, we compare the amount of secure key storage required per FPGA in our proposed framework (combining KAC with BYOK) against a framework that simply uses BYOK. The latter scheme would require to store the AES-128 key for every tenant on the corresponding FPGA partition allocated to her. Naturally, the storage requirement grows with the number of partitions that a given FPGA can support. In our proposition, the aggregation capability of KAC ensures that the tamper-resistant non-volatile storage requirement is independent of number of partitions that a given FPGA can support. In other words, our FPGA provisioning scheme has a far superior scalability in terms of secure key storage, as compared to a simple BYOK-based provisioning scheme.
\item\textbf{On-Chip Resource Overhead:} Since our framework requires only a single KAC decryption engine per FPGA, the on-chip resource overhead remains almost constant with respect to the number of partitions that a given FPGA can support. This is illustrated in Figure~\ref{plot:area}. The only slight increase is due to the presence of an AES-decryption engine in every FPGA partition. However, as demonstrated in Table~\ref{tab:decrypt}, the resource overhead for an AES-128 decryption engine is negligible as compared to the KAC decryption engine. Thus our framework is also scalable with respect to its on-chip resource overhead.
\item\textbf{Bitstream Encryption/Decryption Performance:} Finally, as already mentioned, the bitstream encryption/decryption latency (both KAC and AES-128) of our framework is independent of the number of partitions that a given FPGA can support.  
\end{itemize}
\noindent In summary, the incorporation of KAC plays a crucial role in ensuring that our framework retains the same levels of performance and efficiency for arbitrarily large number of tenants sharing a single FPGA node. To the best of our knowledge, this is the first FPGA provisioning framework to be prototyped that achieves a desirable balance between security and scalability in the multi-tenancy setting.

\section{Conclusion}

In this paper, we proposed a dedicated and scalable framework for secure multi-tenant FPGA provisioning on the cloud. Our framework guarantees bitstream confidentiality in exchange for a constant amount of resource/memory overhead, \emph{irrespective of the number of tenants sharing  a given FPGA}. We achieved this using a novel technique known as \emph{key-aggregation} that is provably secure under well-studied cryptographic assumptions. Our framework can be easily integrated into existing cloud-based infrastructures such as OpenStack, and does not interfere with other desirable properties of an FPGA provisioning mechanism, such as resource virtualization/isolation and platform compatibility. We illustrated the scalability of our proposed approach via a prototype implementation on Xilinx Virtex-7 UltraScale FPGAs. Our results indicate that the proposed approach has a fixed overhead of around $10-15\%$ of the available FPGA resources. This overhead remains unaltered for any number of tenants/workloads using the FPGA resources at any given point of time.

\section{ACKNOWLEDGEMENTS}
The authors would like to acknowledge Intel Corporation, Intel Labs for partial funding of the work under the project ''LightCrypto: Ultra-Light-weight Robust Crypto-Architectures for Performance and Energy".

\bibliographystyle{ACM-Reference-Format}
\bibliography{sample-bibliography} 

\end{document}